# Charge-density depinning at metal contacts of graphene field-effect transistors

Ryo Nouchi[1,a)] and Katsumi Tanigaki[1,2]

[1] *WPI Advanced Institute for Materials Research, Tohoku University, Sendai 980-8578, Japan*

[2] *Department of Physics, Graduate School of Science, Tohoku University, Sendai 980-8578, Japan*

An anomalous distortion is often observed in the transfer characteristics of graphene field-effect transistors. We fabricate graphene transistors with ferromagnetic metal electrodes, which reproducibly display distorted transfer characteristics, and show that the distortion is caused by metal-graphene contacts with no charge-density pinning effect. The pinning effect, where the gate voltage cannot tune the charge density of graphene at the metal electrodes, has been experimentally observed; however, a pinning-free interface is achieved with easily-oxidizable metals. The distortion should be a serious problem for flexible electronic devices with graphene.

[a)] Author to whom correspondence should be addressed; electronic mail: nouchi@sspns.phys.tohoku.ac.jp

Graphene, a one-atom-thick carbon sheet with a honeycomb structure, has attracted significant attention due to its unique physical properties[1] and applicability to future high-speed/spintronics devices. A field-effect transistor (FET) schematically shown in Fig. 1(a), which is a model electronic device, has been utilized to investigate electronic transport properties of graphene. An electric current through graphene is injected/extracted from metallic electrodes (source/drain electrodes in the FET structure), and various kinds of metals has been employed as the electrode materials. Although transfer characteristics (gate voltage $V_G$ dependence of the drain current $I_D$) of graphene FETs typically display a V-shaped curve, there are several reports which show anomalously-distorted transfer characteristics.[2,3] Such distortion indicates a deterioration of gate-voltage response of the device, i.e. a decrease in field-effect mobilities, and thus it is of technological importance to elucidate the origin of the distortion.

However, the distortion has not been paid so much attention to date. The less attention might be due to the fact that a heat treatment in inert atmosphere, which has been originally developed for the removal of on-graphene impurities such as resist residues,[4] can eliminate the distortion. To address this question, graphene transistors were fabricated by mechanical exfoliation and conventional electron-beam lithography procedures. Figure 1(b) shows transfer characteristics of a graphene FET with Ni source/drain electrodes before (dashed line) and after (solid line) annealing at 400 °C in Ar atmosphere. The distortion is visible in the negatively-gated region of the characteristics measured just after the device fabrication [Fig. 1(b); dashed line], but it disappeared after the annealing procedure [Fig. 1(b); solid line]. From the fact that the annealing procedure in order to remove the impurities on graphene (such as resist residues) can eliminate the distortion, it has been believed that the on-graphene impurity is the origin of the distortion. However, it may be too naive to accept the conclusion because the impurity scenario is unaccountable for the fact that graphene FETs with Au source/drain electrodes do not show such distortion even if the identical fabrication procedure is employed.[2]

As is discussed above, the origin of the distortion in transfer characteristics of graphene FETs with source/drain electrodes made of metals other than Au should be different from the on-graphene impurities. Figure 2 shows the transfer characteristics of Ni-contacted graphene FETs with different channel lengths, *L*. All the characteristics display anomalous distortion in the negatively-gated region, as previously reported.[2,3] The devices with shorter channel lengths exhibit larger distortions, which indicates that the distortion is induced by the metal contacts, because shorter channels correspond to smaller channel resistances, and thus the contact-related effect becomes more dominant in the overall device resistance.[2] It has been predicted for metal/graphene-nanoribbon/metal junctions that the mismatch of propagating modes in metal and graphene-nanoribbon regions makes an insulating behavior even at gate voltages far from the Dirac point.[5,6] This can induce several distortions in transfer characteristics; however, in this study micron-scale graphene flakes were employed instead of graphene nanoribbons, and the shapes of the transfer characteristics shown in Fig. 2 are different from those predicted for metal/graphene-nanoribbon/metal junctions.

To investigate the cause of the anomalous distortion observed in the transfer characteristics, an attempt was made to reproduce the distortion by means of a simple transfer characteristics simulation. The starting point is the assumption that the overall channel resistance can be expressed as a series connection of local resistances. If the graphene channel is homogeneous along the direction parallel to the metal contact edges, then the overall resistance $R$ is expressed by considering only the direction from the source electrode edge ($x$=0) to the drain electrode edge ($x$=$L$):

$$R = \frac{1}{W}\int_0^L \rho(x)dx = \frac{1}{W}\int_0^L \frac{1}{\sigma(x)}dx\,, \tag{1}$$

where $W$ is the channel width, $\rho(x)$ is the local resistivity at distance $x$, and $\sigma(x)$ is the local conductivity at $x$. In the Drude model, multiplying the charge density in graphene $ne$ by the charge carrier mobility $\mu$ gives $\sigma(x)$, where $n$ is the number density of charge carriers and $e$ is the elementary charge. From a parallel-plate capacitor model, the charge density is related to the gate voltage through the relationship, $ne=C_0V_G=\varepsilon_0\varepsilon_r V_G/d$, where $C_0$ is the gate capacitance per unit area, $\varepsilon_0$ is the permittivity of a vacuum, $\varepsilon_r$ is the relative permittivity of the gate dielectric (3.9 for $SiO_2$), and $d$ is the thickness of the gate dielectric (300 nm in this study). However, a typical $V_G$ dependence of the conductivity for graphene indicates one typical feature; the conductivity remains finite ($\sigma_{min}$) even if $V_G$ is at the Dirac point $V_D$, where the carrier density is zero. To mimic this feature, a phenomenological expression is introduced as

$$\sigma(x)=\sqrt{\left\{\mu\frac{\varepsilon_0\varepsilon_r}{d}\left(V_G-V_D(x)\right)\right\}^2+\sigma_{min}{}^2}\ . \tag{2}$$

This local conductivity equals $\sigma_{min}$ at the local Dirac point [$V_G=V_D(x)$], and it is asymptotic to the Drude value $\mu\varepsilon_0\varepsilon_r\{V_G-V_D(x)\}/d$, with high $V_G$s. $R$ can then be rewritten as

$$R=\frac{1}{W}\int_0^L\left[\left\{\mu\frac{\varepsilon_0\varepsilon_r}{d}\left(V_G-V_D(x)\right)\right\}^2+\sigma_{min}{}^2\right]^{-1/2}dx\ . \tag{3}$$

The conductivity of the entire channel $\sigma_{total}$ is obtained as $L/(RW)$.

To simulate the transfer characteristics, the local doping profile $V(x)\equiv V_G-V_D(x)$, which represents the local doping level including the contributions of carrier doping from the metal electrodes and electrostatic doping by the gate electric field, and its gate-voltage response should be known. At metal-graphene interfaces, carriers are also expected to be doped from the metal,[7] and the resulting band bending has been observed by means of scanning photocurrent microscopy (SPCM).[8,9] Figure 1(c) shows an assumed doping profile for the Ni-contacted graphene FETs. Linearly graded doping takes place from the contact edges to the distance $L_d$, and with $V_G$=0 the doping level becomes zero at points distant from the contact (from $x=L_d$ to $x=L-L_d$). In this profile, $V(x)$ with zero $V_G$ near the contacts has positive values, which indicates electron doping from the metal electrodes to the graphene channel near the electrode edges. The SPCM measurements[8,9] also showed that the electrostatic potential close to Au contacts does not change by application of gate voltages, which indicates that gate voltages cannot be used to effectively tune the charge density near the Au contacts. If the same holds for the graphene FETs in this study, the gate-voltage dependence of the doping profile shown in the leftmost panel of Fig. 3(a) can be assumed. However, the simulated transfer characteristics in Fig. 3(a) do not reproduce the distortion, although they do indicate an asymmetry between the positively- and negatively-gated regions, as previously discussed in the context of doping from metal contacts.[10]

Figure 3(b) shows the simulated transfer characteristics without the charge-density pinning effect at the metal contacts. As displayed in the leftmost panel of Fig. 3(b), the doping level at the metal contacts can also be effectively tuned by application of gate voltages, similar to the central region of the channel. The simulated characteristics clearly show the distortion in the negatively-gated regions, and the distortion is more apparent with shorter channels. These are the same features that were experimentally observed with the Ni-contacted devices (Fig. 2), which suggests that charge-density *depinning* occurs at Ni contacts, unlike the pinned charge density at Au contacts.[8,9]

One probable explanation for the depinning of the charge density near metal-graphene contacts is oxidation of the metal surfaces. The pinning effect, i.e., the gate-uncontrollable charge density at the metal contacts, may be attributed to the screening of the gate electric field at the contacts. Electrons are known to spill out from the metal surface and the electron density becomes finite, even outside of the metal.[11] The electron density outside the metal is lower than that inside the metal, but it may still be sufficiently high to screen out the external electric field. However, if a thin metal oxide layer is present at the metal-graphene interface, then the spilled electron density largely decays inside the oxide layer. The resultant low electron density inside the graphene is no longer able to screen out the gate electric field. It should be noted that the spilled electron discussed here is different from the carriers doped from metal contacts which lead to band bending shown in Figs. 1(c) and 3.

Metal oxide formation at the interface with graphene is necessary to weaken the charge-density pinning effect. Metal electrodes were formed by the direct deposition of metal atoms onto graphene, and the metal atoms at the interface seem to be guarded against oxidation. However, it was reported that an Al electrode directly deposited onto a thin graphite film behaves as a gate electrode without intentional insertion of an insulating layer.[12] This observation indicates that Al electrodes can be considerably oxidized even at the Al-graphene interface, which means the unintentional formation of a gate dielectric at the interface. It was also found that sufficient insulation is quickly recovered in air after the insulation breaks down due to overpotential at the Al electrode.[12] These results strongly suggest that oxygen can penetrate through the metal-graphene contacts, so that the metal surface can oxidize at the interface, which should happen with relatively reactive metals. Ferromagnetic metals are more reactive than Au and are easily oxidized. Charge density pinning should occur at the Au-graphene contacts as experimentally observed,[8,9] but the charge density can also be depinned at reactive metal contacts as discussed in this study.

The effect of air on the transfer characteristics of a ferromagnetically-contacted device was examined to confirm this scenario. Figure 3(b) shows the transfer characteristics measured immediately after device fabrication (dashed line) and after exposure to air for 125 h (solid line). The transfer characteristics clearly display a large distortion in the negatively-gated region after exposure to air, which would induce further oxidation of the metal electrodes. The scenario predicts that the development of oxidation at the metal-graphene interfaces causes weaker screening of the gate electric field, i.e., weaker pinning of the charge density at the metal contacts. The weaker pinning manifests as a larger distortion of the transfer characteristics. The results displayed in Fig. 3(b) strongly support oxidation-induced depinning of the charge density.

The shape of the simulated transfer characteristics can be understood by analogy with double-gated graphene FETs. In a double-gated device, the top gate is usually formed on part of the graphene channel, whereas the back gate covers the entire channel region.[13] The top gate modulates the Fermi level of the channel only underneath the top gate, which means that $V_D(x)$ of the top-gated region is different from that of the rest of the channel. If such a condition is achieved by application of a finite top-gate voltage, then the transfer characteristics exhibit two local minima corresponding to the Dirac points of the top-gated region and the remaining region. The distortion characteristics reproduced in this study can be understood by considering two distinct regions, as with a double-gated device; namely, the region doped from the metal contacts, and the undoped region. Thus, the gate voltage at the kink in the negatively-gated region corresponds to the Dirac points of the contact-doped region, $V_D(0)$ and $V_D(L)$. In the simulation results shown in Fig. 3, the value of $V_D(0)$ [$=V_D(L)$] was chosen to reproduce the experimentally-observed features of the transfer characteristics and was set at −60 V for the Ni-contacted devices. This value corresponds to a Fermi level shift of −0.24 eV.

In conclusion, the electronic transport properties of graphene were investigated experimentally using a FET structure with Ni source/drain electrodes. A distortion induced by the Ni contacts was observed in the negatively-gated regions of the transfer characteristics of the Ni-contacted FETs. The distortion was reproduced with a series-resistance model by excluding charge-density pinning at the metal contacts. The pinning-free interface is contrary to that which has been experimentally observed to date.[8,9] A possible mechanism for the depinning is oxidation of the metal surfaces at the metal-graphene interfaces, which is supported by the increase in distortion after exposure to air. Such distortion is indicative of the potent influence of metal-to-graphene carrier doping on the electronic transport properties of graphene. While strong pinning induces subtle asymmetry in the transfer characteristics,[10] no (or very weak) pinning results in a large distortion of the transfer characteristics. The present study has elucidated the significance of the metal-graphene interfacial conditions. In addition, the metal-contact induced distortion must be of significant importance to flexible electronic devices with graphene, where a flexible plastic substrate is used instead of a rigid $SiO_2$/Si substrate. Graphene is known as one of strongest materials in terms of Young’s modulus and elastic stiffness[14] and is also very flexible because of its two-dimensional structure, which indicates that graphene is suited to the flexible electronics. However, the plastic substrate necessary for flexible devices is weak against a heat treatment which sometimes eliminates the distortion. Our investigations suggest that inert metallic materials such as Au should be employed to construct flexible electronic devices with graphene.

The authors are grateful to S. Okada for valuable discussions, and M. Murakami and M. Shiraishi for supplying the graphite crystal. This work was supported in part by the Foundation Advanced Technology Institute.

Figures

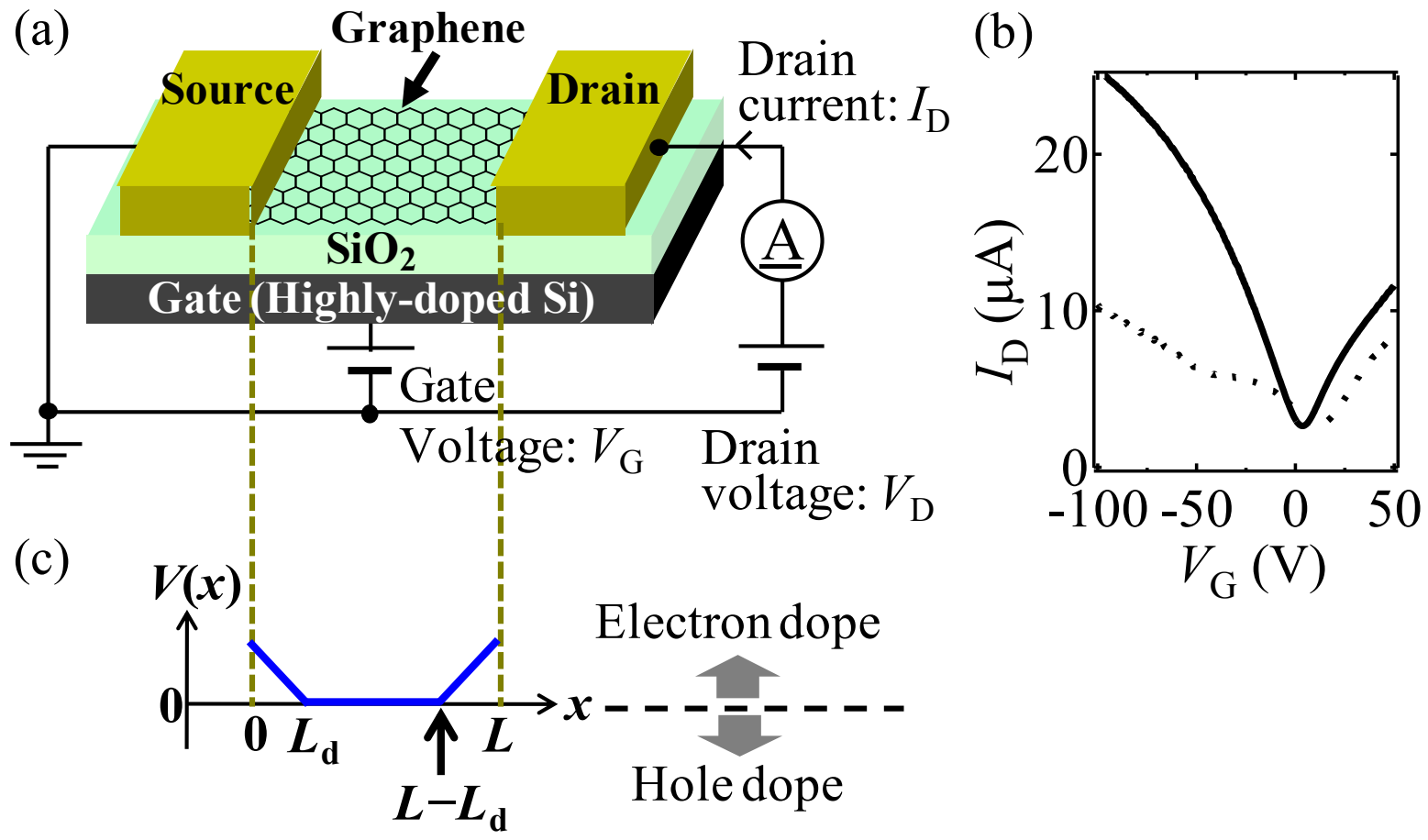

FIG. 1. (Color online) (a) Schematic diagram of the graphene field-effect transistor fabricated in this study. (b) Transfer characteristics of a Ni-contacted graphene FET measured before (dashed line) and after annealing at 400 °C in Ar atmosphere for 12 h (solid line). The channel length of the device is 7.8 μm. The drain voltage was set to 0.1 V. (c) Assumed doping profile along the graphene channel without application of gate voltages. The vertical axis is a gate-voltage equivalent and positive (negative) values correspond to electron (hole) doping. Linearly graded doping takes place from the source/drain electrode edges to the distance $L_d$, and the doping level becomes zero at points distant from the contact ($L_d \le x \le L - L_d$).

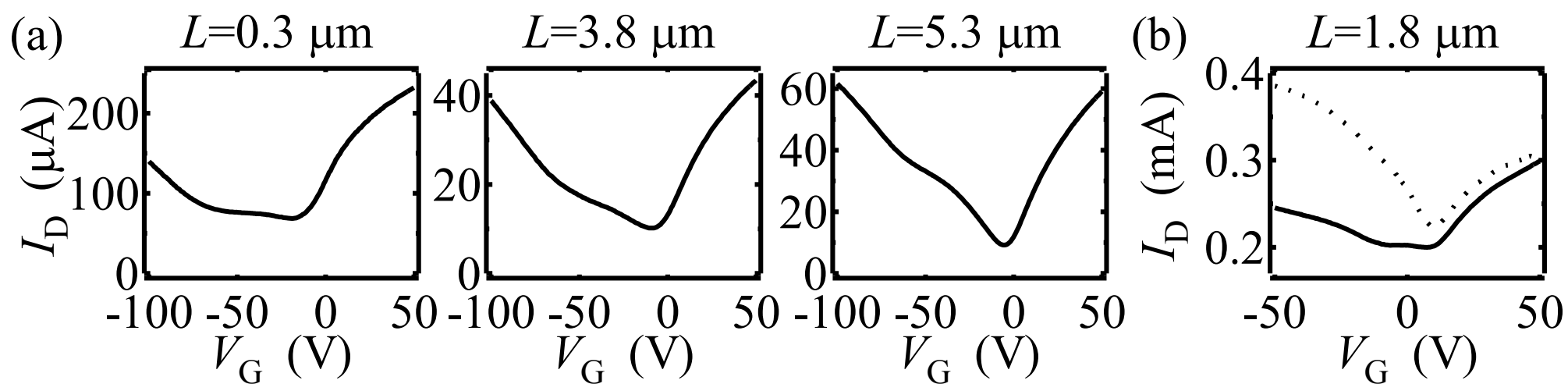

FIG. 2. Transfer characteristics of (a) Ni-contacted field-effect transistors with different channel lengths, $L$, and (b) a Co-contacted FET measured immediately after device fabrication (dashed line) and after exposure to air for 125 h (solid line). The drain voltage was set to 0.1 V.

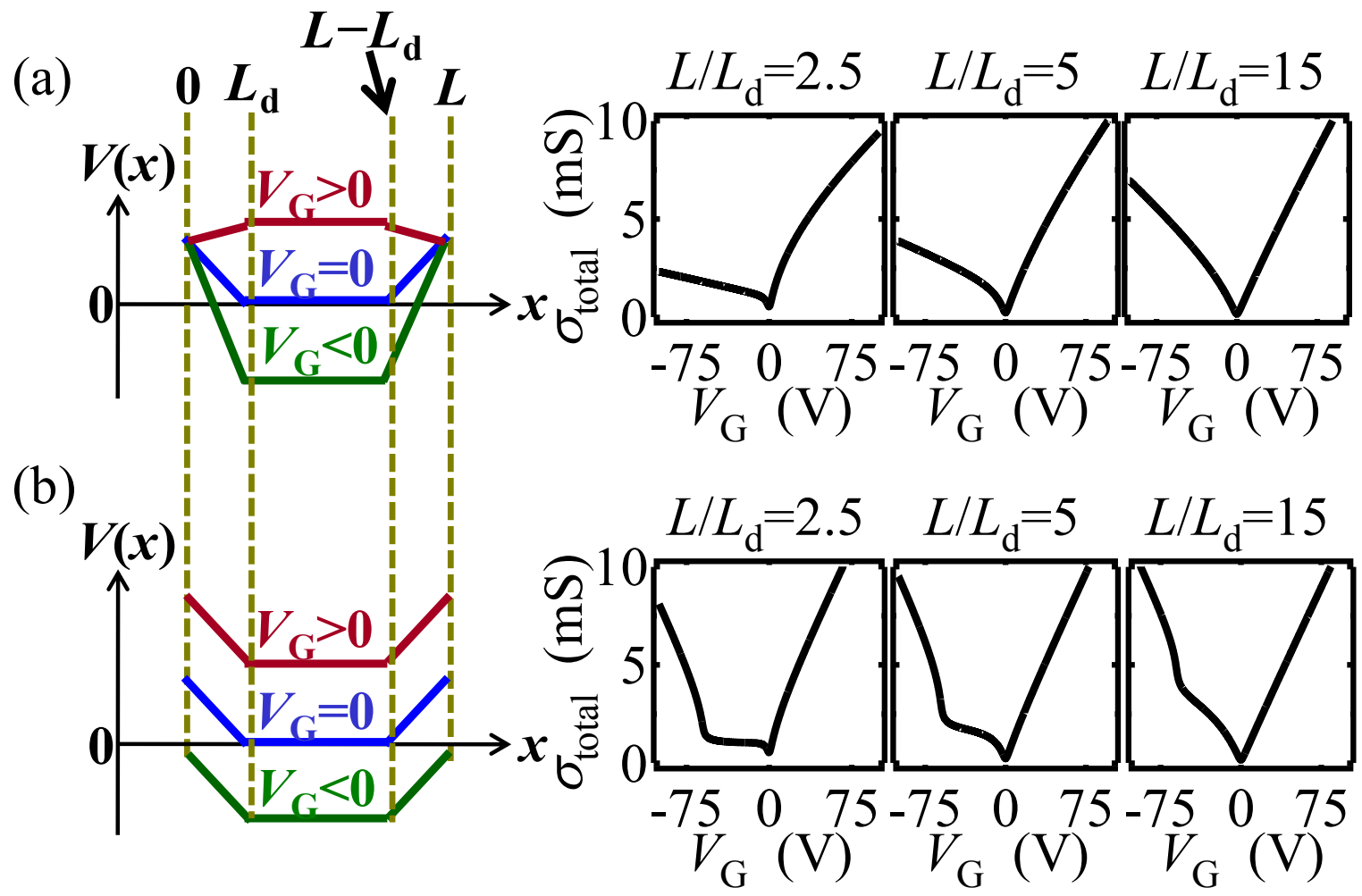

FIG. 3. (Color online) Simulated transfer characteristics of graphene field-effect transistors with different channel lengths, $L$. The leftmost panels are the assumed gate-voltage dependences of the doping profiles (a) with and (b) without charge density pinning at the metal contacts. Without such a pinning effect, the doping level near the metal electrodes can be effectively tuned by application of gate voltages, similar to the central region of the channel.